\documentclass[preprint]{aastex}

\begin{document}

\title{ROSAT HRI and ASCA Observations of the Spiral Galaxy NGC 6946 and Its Northeast Complex of Luminous SNRs}
\author{Eric M. Schlegel\altaffilmark{1}, William P. Blair\altaffilmark{2}, \& Robert A. Fesen\altaffilmark{3}}
\altaffiltext{1}{Smithsonian Astrophysical Observatory, 60 Garden Street, Cambridge, MA  02138}
\altaffiltext{2}{Department of Physics and Astronomy, Johns Hopkins University, 3400 N. Charles St., Baltimore, MD  21218}
\altaffiltext{3}{Department of Physics and Astronomy, 6127 Wilder Lab, Dartmouth College, Hanover, NH, 03755}

\begin{abstract}

Analysis of 80 ksec {\it ASCA} and 60 ksec {\it ROSAT} HRI
observations of the face-on spiral galaxy NGC 6946 are presented.  The
{\it ASCA} image is the first observation of this galaxy above $\sim$2
keV.  Diffuse emission may be present in the inner $\sim$4$'$
extending to energies above $\sim$2--3 keV.  In the HRI data, fourteen
point-like sources are detected, the brightest two being a source very
close to the nucleus and a source to the northeast that corresponds to
a luminous complex of interacting supernova remnants (SNRs).  We
detect a point source that lies $\sim$30$''$ west of the SNR complex
but with a luminosity $\sim$1/15 of the SNR complex.  None of the
point sources shows evidence of strong variability; weak variability
would escape our detection.

The {\it ASCA} spectrum of the SNR complex shows evidence for an
emission line at $\sim$0.9 keV which could be either Ne IX at
$\sim$0.915 keV or a blend of ion stages of Fe L-shell emission if the
continuum is fit with a power law.  However, a two component,
Raymond-Smith thermal spectrum with no lines gives an equally valid
continuum fit and may be more physically plausible given the observed
spectrum below 3 keV.  Adopting this latter model, we derive a density
for the SNR complex of 10--35 cm$^{-3}$, consistent with estimates
inferred from optical emission line ratios.  The complex's extraordinary
X-ray luminosity may be related more to the high density of the
surrounding medium than to a small but intense interaction region
where two of the complex's SNRs are apparently colliding.
\end{abstract}


\section{Introduction}

The first step in the study of the X-ray emission from spiral galaxies
is the identification and characterization of a galaxy's point and
extended sources, followed by a comparison of these source types
across a broad range of galaxy properties.  Prior to the launch of
{\it ROSAT} in 1990, X-ray observations of spiral galaxies were
limited to Local Group members or a few bright and relatively nearby
galaxies such as M51, M81, and M83 (e.g., \citet{F92} and references
therein).  {\it ROSAT} greatly expanded the number of galaxies for
which more detailed observations could be undertaken due in part to a
sharper point spread function and a lower detector background
\citep{Pf86}.

X-ray emission from the face-on, nearby, spiral galaxy NGC 6946
(d=5.1Mpc, \citealt{deV79}; N$_{\rm H}$ $\sim$2--3$\times$10$^{21}$
cm$^{-2}$, \citealt{BH84}) was first observed with {\it Einstein}
\citep{FT87}. However, the relatively low spatial resolution of the
data was insufficient to permit individual source identifications.
More recent {\it ROSAT} PSPC observations detected about 10 sources
\citep{S94a}, two of which have been identified as counterparts to
optically-detected supernova remnants (SNRs) \citep{S94b, BF94, VD94,
MF97}.  One of these is of particular interest due to its unusually
high optical and X-ray luminosity ($\sim$10$^{40}$ erg s$^{-1}$ in the
0.2--2.0 keV band).  Despite unusually high optical, radio, and X-ray
luminosities suggestive of a young SNR, optical spectral data showed
only relatively low expansion velocities.  A {\it Hubble Space
Telescope} observation of this source has revealed that it is actually
a complex of two or more unusually luminous remnants with an
especially bright region marking an apparently strong collision
between two of the remnants \citep{BFS00} (hereafter BFS00).  The NGC
6946 SNR complex may be a younger example of the double SNR DEM L316
in the LMC \citep{Wil97}.  \citet{DGC00} present an alternative
interpretation, describing MF16 as the interaction of a strong wind
with a nitrogen-rich circumstellar medium.

In addition to the X-ray point sources, unresolved emission was also
detected across much of the galaxy in the PSPC data \citep{S94a}.
This emission was interpreted as arising from a diffuse component
likely representing emission from the galaxy's hot interstellar
medium.

In this paper, we present both the {\it ROSAT} HRI image and an {\it
ASCA} CCD observation \citep{TIH94} of NGC 6946.  These data consist
of images of the galaxy and a spectrum of the extraordinary SNR,
hereafter called MF16 (the 16th entry on the SNR list of
\citealt{MF97}).

\section{Observations} \label{SecObs}

\subsection{{\it ROSAT} HRI and PSPC}

The {\it ROSAT} HRI was used to observe NGC 6946 for 60.3 ksec spread
across 15 days in 1994 May (Table~\ref{sum_Xobs}).  We used the HRI
analysis software assembled by S. Snowden \citep{Sn98} to remove the
particle background from the data, leaving 59.6 ksec of
deadtime-corrected exposure.  Figure~\ref{HRI_fig} shows the results
after filtering and smoothing with an pyramid-shaped, adaptive filter
\citep{Lor93} of 10 counts.  The 10-count criterion represents a
compromise between the preservation of spatial resolution and
accumulating a high signal-to-noise ratio across the entire image.
The HRI data have been overlaid onto an optical image from the
digitized Palomar Sky Survey in Figure~\ref{HRI_opt}.

Point sources were detected using the {\it Chandra} Observatory's {\it
wavdetect} program, based on the wavelet code of \citet{Mic96}.  We
used a detection criterion of 1.0$\times$10$^{-6}$ which permits at
most $\sim$1 false source.  Point sources have been labeled and
correspond to the entries in Table~\ref{pt_srcs}.  The luminosities in
Table~\ref{pt_srcs} are tabulated for the 0.5-2.0 keV band assuming a
1 keV thermal bremsstrahlung spectrum absorbed by a Galactic column
N$_{\rm H}$ of 2.5$\times$10$^{21}$ cm$^{-2}$.  The luminosities
differ from the adopted model by $\sim$25\% if the adopted model
temperature is instead set to 5-7 keV.

Since the {\it ROSAT} PSPC data were published \citep{S94a}, several
improvements in the pipeline processing have been made, particularly
in the corrections for the boresight and aspect \citep{Brl96}.  The
data were obtained in 1992 June (Table~\ref{sum_Xobs}).  The
re-processed data were filtered using the diffuse analysis procedures
available from the {\it ROSAT} data center and described by
\citet{Sn95}.  After screening the data for particle background and
solar-scattered X-rays, a total of 34.3 ksec remained.  The filtered
PSPC image does not differ in any systematic way from that presented
in \citet{S94a} so we do not include it to conserve space.

\subsection{{\it ASCA}}

The {\it ASCA} data were obtained from 1994 Dec 28 to 1995 Jan 2 for a
total exposure time of 92 ksec (Table~\ref{sum_Xobs}).  After applying
the standard screening 83 ksec of observation remained.  The majority
of the data ($\sim$75\%) were obtained in 1-CCD mode with the
remainder taken in 2-CCD mode.  Data screening was applied to each
data mode separately and the results were registered and summed.  The
data were divided by energy into the following seven bands: 0.5--1.0,
1.0--1.5, 1.5--2.5, 2.5--3.5, 3.5--4.5, 4.5--6.5, and 6.5--9.5 keV.  The
corners of each image were used to define the background level which
was then subtracted.

The point spread function (PSF) of the {\it ASCA} mirrors has a FWHM
of $\sim$3$'$ and has a cross-shaped pattern \citep{TIH94}.  To
recover the most information from the image of NGC 6946, we
deconvolved the PSF from the image using a point source filtered into
the seven energy bands and the Lucy-Richardson deconvolution (the task
as coded in the IRAF/STSDAS\footnote{IRAF is distributed by the
National Optical Astronomy Observatories which is operated by the
Associated Universities for Research in Astronomy, Inc. under
cooperative agreement with the National Science Foundation.  The Space
Telescope Science Data Analysis System is distributed by the Space
Telescope Science Institute.} ``restore'' package) following the
prescription in \citet{JGZ93}.  The noise in the corners of the image
were monitored by calculating the root mean square average value after
each iteration.  The restoration was stopped when the rms value
approached an asymptotic value (typically, $\sim$20 iterations).

The relatively high foreground column density to NGC 6946 nearly
eliminates any photons below 0.5 keV, so we did not include any counts
below that energy.  Figure~\ref{ASCA_montage} shows a montage of the
results; the energy band increases left to right, top to bottom.  

As a test of the deconvolution, we blurred the {\it ROSAT} HRI data
with the {\it ASCA} PSF.  To obtain the PSF as a function of energy,
we filtered the data of a point source (3C273, sequence 70023000) into
the seven {\it ASCA} energy bands defined above.  We then convolved
those PSF bands with energies $<$2.5 keV with the HRI image and summed
the results.  We also filtered the raw {\it ASCA} counts to retain
only those events with E $<$2.5 keV.  Figure~\ref{hri_blur} shows the
summed image and the filtered raw {\it ASCA} data.  Note the
difference between the real (lower right) and the artificial (upper
right) {\it ASCA} data.  Figure~\ref{hri_blur} will be used below to
help interpret the deconvolved {\it ASCA} data.

\section{Analysis and Discussion} \label{SecAnalysis}

\subsection{{\it ROSAT} HRI Data} \label{SecAnal_HRI}

The {\it ROSAT} HRI image of NGC 6946 was used to inventory the
galaxy's point sources.  Since the {\it ROSAT} HRI offers little
spectral resolution \citep{Pr97}, we used a sliding box detection
algorithm (box size 8$''$) to locate and to identify candidate point
sources.  Table~\ref{pt_srcs} lists the sources detected and their
count rates.  HRI sources are designated with an ``H'' followed by a
serial number.  The counts were extracted using circular apertures
48$''$ in diameter.  The background was determined from an aperture
5$'$ in diameter positioned well away from the galaxy and corrected
for vignetting at that location.

The HRI image (Figure~\ref{HRI_fig}) shows that many of the point
sources identified in the PSPC data \citep{S94a} are still present and
remain at essentially identical relative intensities as in the PSPC
data.  In particular, HRI sources H1, H3, H5, H6, H9, H10 (SNR MF16),
and H11 are all clearly detected in the PSPC image.
Table~\ref{pt_srcs} lists the correspondences between the sources
detected in the HRI data and those detected in the PSPC data.
Spectral fits and hardness values from the PSPC data showed that all
of the sources within 8$'$ of NGC 6946 were consistent with sources in
NGC 6946 (or behind it) \citep{S94a}.  Based on a Galactic log N-log
S relation \citep{Kra99}, we expect $\sim$0.1 stars to fall by chance
within a radius of 5$'$ of NGC 6946.  Using their extragalactic log
N-log S (their ``AGN'' curve), we expect 0.04 AGN within the same 5$'$
radius.

If we correct for the count rate differences between the PSPC and the
HRI, all of the sources are consistent with a constant between the
PSPC and HRI epochs.  The large uncertainty in the conversion from
PSPC to HRI counts prohibits a stronger statment.  Within the HRI
data, individual pointings are less than $\sim$2 ksec in length.  Only
the strongest sources (H5, H9, H10) were examined for variability and
none show evidence of variability $>$90\% confidence.  The X-ray
constancy of sources in NGC 6946 is in marked contrast to some other
galaxies, for example, NGC 1313, where about a half-dozen sources have
been detected as variable objects \citep{Sch+99}.  We turn now to a
discussion of the individual sources.

The extraordinary X-ray luminosity of the MF16 SNR complex is obvious
in Figs. 3--5 where it is the dominant X-ray source in NGC~6946.
Therefore its X-ray spectral properties are of special interest.  The
MF16 complex and environs are resolved in the HRI data into two
sources; MF16 itself (H10) and a fainter additional source, H8,
$\sim$30$''$ farther to the west-southwest (see Fig. 1) with a
luminosity $\simeq$ 6.5\% that of H10.  H8 produces significant count
contamination in the MF16 spectrum in both the {\it ROSAT} PSPC and
{\it ASCA} data, leading to uncertainty in the spectral behavior of
MF16.

Unfortunately, the nature of H8 is unknown.  It is not directly
attributable to an OB association that is near MF16 (BFS00) because H8
lies ${\sim}10''$ west of the association. The luminosity of H8 is
$\sim$10$^{38}$ erg s$^{-1}$ and thus not unlike that of X-ray
binaries \citep{WNP95}, suggesting one possible identification. Its
contamination of H10 (MF16) is particularly serious if H8 has a
systematically different spectrum than the SNR complex.  As an
example, if MF16 can be described by a Raymond-Smith plasma at 1 keV
and the putative XRB by a thermal bremsstrahlung component of 5 keV,
as is typical of X-ray binaries, the merged spectrum will be
increasingly dominated by the 5 keV bremsstrahlung emission above
$\sim$3 keV even with a 15:1 ratio in the total 0.5--2.0 keV
luminosities.  This will make the interpretation of the {\it ASCA} SIS
data more difficult.

The source H5, although close to the center of NGC 6946, may not
correspond to the X-ray emission from the galaxy's nucleus.  The
difference in the position of MF16 between the HRI data and the VLA
data \citep{VD94} amounts to ${\Delta}{\alpha}$ = +0$''$.6,
${\Delta}{\delta}$ = --1$''$ (defined as HRI -- VLA).  Only one other
source in the HRI field is positively identified: the Algol variable
DT Cep which lies 12$'$ off-axis.  The difference between the optical
and HRI positions are ${\Delta}{\alpha}$ = +6$''$, ${\Delta}{\delta}$
= --9$''$ (defined as HRI -- optical) \citep{Kra99}.  These
differences are typical of the uncertainties of the {\it ROSAT}
boresight offsets \citep{KH93}.  If we use these MF16 offsets to
correct the position of H5, because they are nearly on-axis, the
revised X-ray coordinates (Table~\ref{pt_srcs}) place it within
4--5$''$ of the radio nucleus\footnote{The position of the radio
nucleus is (J2000) ${\alpha}$: 20:34:52.33; ${\delta}$: +60:09:14.23
\citep{VD94}.}.  In the optical, the center of light, assumed to be
the nucleus, appears to lie {\it south} of the position of the radio
nucleus and H5 by $\sim$15--20$''$.  The PSPC spectrum \citep{S94a} of
the ``nuclear'' source is soft with a fitted thermal bremsstrahlung
temperature of $\sim$0.5 keV.  The source is consistent with the point
spread function of the HRI.  The 0.5--2.0 keV luminosity is
$\sim$3$\times$10$^{38}$ erg s$^{-1}$ which is highly sub-Eddington
($\sim$3$\times$10$^{-7}$) if the nucleus were considered to be an
accreting black hole of $\sim$10$^7$ M$_{\odot}$ based on the black
hole mass-bulge luminosity correlation (e.g., \citealt{KR95}) and the
bulge luminosity of $\sim$-17.1 \citep{SdV86,KC84}.

Recently, \citet{CM99} have shown that a population of off-set
``nuclear'' sources may exist at the center of galaxies.  They found
that, of those nearby (v$_{\rm red}$ $<$ 1000 km s$^{-1}$) galaxies
for which X-ray observations exist, about half contain an X-ray source
consistent with an off-center point source, each of which has a
luminosity of $\sim$10$^{39-40}$ erg s$^{-1}$.  They suggest these
sources are 10$^{2-4}$ M$_{\odot}$ black holes.  NGC 6946 was one of
the galaxies on their list of detections.

Four sources are detected in the HRI image that are weak, but visible
in the PSPC data.  These are H2, H12, H13, and H14.  PSPC source 5,
which lies $\sim$1$'$.5 northwest of MF16 and has an L$_{\rm X}$ of
5.5$\times$10$^{37}$ erg s$^{-1}$, is not detected in the HRI
observation.  A drop even as small as $\sim$30 in its luminosity would
place the source below the HRI detection threshold.  Source H7 was not
resolved in the PSPC data.  In that data set, it sits within the broad
band of unresolved emission near the nucleus of NGC 6946.

In the HRI data, we detect none of the diffuse emission seen in the
PSPC data, but we can easily understand why.  The conversion factor of
HRI counts to PSPC counts ranges from $\sim$2.2 \citep{S96} to
$\sim$2.8 \citep{Zim94} depending upon the adopted spectral shape.  We
arbitrarily adopt a conversion of 2.5 HRI counts to 1.0 PSPC count so
we can illustrate the limitations of the HRI for the detection of
diffuse emission.  To reach a sensitivity similar to that of the PSPC,
the HRI exposure time must be longer than the PSPC exposure time by at
least the conversion factor.  Strictly on the basis of the conversion
factor, the 60 ksec HRI exposure corresponds to an approximate PSPC
exposure of 24 ksec.  This is substantially less than the actual
duration of the PSPC observation in which the diffuse emission was
detected \citep{S94a}.  An alternative approach illustrates the same
point.  The PSPC-detected surface brightness is
$\sim$5$\times$10$^{36}$ erg s$^{-1}$ arcmin$^{-2}$ (value from
\citet{S94a} corrected for arithmetic error).  That brightness
generates approximately 10$^{-4}$ counts s$^{-1}$ arcmin$^{-2}$ in the
PSPC while the internal PSPC background is $\sim$3$\times$10$^{-5}$
cts s$^{-1}$ arcmin$^{-2}$ \citep{Snow92}, for a ratio of $\sim$3.
For the HRI, the same surface brightness produces
$\sim$6$\times$10$^{-6}$ counts s$^{-1}$ arcmin$^{-2}$ compared to the
internal background of $\sim$4$\times$10$^{-3}$ counts s$^{-1}$
arcmin$^{-2}$ for a ratio of $\sim$0.002 \citep{Dav95}.  The diffuse
emission is overwhelmed by the internal background of the HRI.

\subsection{{\it ASCA} Data}

The deconvolved {\it ASCA} image of NGC 6946 is shown in
Figure~\ref{ASCA_montage}.  This is the first image of NGC 6946 at
energies above 2 keV.  We filtered the {\it Einstein} IPC image into E
$<$2 keV and $>$2 keV; there are no photons above $\sim$2 keV.  The
{\it ASCA} can be compared to the HRI image blurred with the {\it
ASCA} PSF shown in Figure~\ref{hri_blur}.  Unlike the blurred HRI data
which contains no true diffuse emission, the {\it ASCA} image shows
point sources which are embedded in diffuse emission.  Below we will
use the the HRI image shown in Figure~\ref{hri_blur} to interpret the
deconvolved {\it ASCA} data.

The bright point sources (H3, H5, H9, and H10) are deconvolved
(although the deconvolution does not separate H5 and H9 except in the
1.0--1.5 keV panel where the counts are highest).  Diffuse emission
surrounds the nuclear sources, most visibly present in the 1--2 keV
range (additional discussion follows below).  There is also crude
spectral information visible in the deconvolved figure for several of
the point sources.  Sources H3, H6, and H9 are weak above $\sim$3 keV.
For H5 and H9, the centroid of emission shifts from H9 to H5 between
the 0.5--1.0 keV panel (upper left) and the 4.5--9.5 keV panel (center
row, right), implying that $\sim$30\% of the flux of H5 lies above 3
keV.  This behavior matches the spectral fit to the PSPC data of H5
which assigned only a lower limit to the temperature of $\gtrsim$0.5
keV \citep{S94a}.

The deconvolved images also provide support for the identification of
H8 as an X-ray binary.  Even though the binary is blended with MF16,
the shape of the deconvolved image becomes an ellipse above $\sim$5
keV, oriented east-west.  We used the `ellipse' task in the STSDAS
package\footnote{} to fit ellipses to the images in the 1.0-2.5 and
4.5-9.5 keV bands.  The ellipticity in the higher-energy panel is
significant at the $\sim$4${\sigma}$ level, while the lower energy
fits are consistent with zero ellipticity, both outwards to an
intensity level 70\% below the maximum.  That implies either
MF16 is contributing a decreasing fraction of the flux of the total
flux or H8 contributes an increasing fraction.  If H8's spectrum is
typical of our Galaxy's population of low-mass XRBs, we expect to see
a bremsstrahlung spectrum with kT $\sim$7 keV in the {\it ASCA} band
(e.g., \citealt{WNP95}).  For such a spectrum, normalized to the count
rate in the HRI image, we estimate $\sim$60\% of the flux to lie above
3 keV and $\sim$25\% above 6 keV.  These estimates increase if H8 is
more similar to a Galactic high-mass X-ray binary.  The deconvolved
{\it ASCA} image suggests H8 is not harder than $\sim$4--7 keV because
otherwise we would see a definite point source in the hardest {\it
ASCA} bands.  An observation with higher spatial resolution and
moderate to good spectral resolution is required to establish
definitively the source type for H8.

Before leaving the discussion of the point source population, we note
the publication of a recent radio survey of NGC 6946 by \citet{LDG97}.
They used the VLA at 6 and 20 cm to carry out a search for compact
radio sources.  Source 85 on their list corresponds to the MF16 SNR
complex to within $\sim$0$''$.5 verifying the identification of
\citet{VD94}.  The correspondence of other sources is included in the
notes of Table~\ref{pt_srcs}, but can be summarized briefly.  Each of
the ``strong'' sources (those with S/N $>$4.5) has a compact radio
counterpart except H1 and H9.

Finally, there is apparent diffuse emission distributed across the
face of the galaxy in the deconvolved images
(Figure~\ref{ASCA_montage}).  This emission was detected in the PSPC
image \citep{S94a}.  If the deconvolution process has recovered all of
the photons of the point sources, then the spectrum of the diffuse
component extends to energies of $\sim$3 keV.  Support for the
correctness of the deconvolution is the match in the overall
appearance of Figure~\ref{ASCA_montage} to the PSPC image \citep{S94a}
where we know diffuse emission was detected.  To examine the
significance of the potential emission, we extracted all counts in an
8$'$ radius circle encompassing the galaxy, excluding the counts
within 2$'$.5 surrounding each of the point sources.  The excluded zone
leaves $\sim$30\% of the flux of a point source outside of the
aperture \citep{JGZ93}.  The counts in the galaxy aperture, in the
0.5-2.0 keV band in common with the PSPC, total
$\sim$5$\times$10$^{-5}$ counts s$^{-1}$ arcmin$^{-2}$.  The
background is $\sim$2.0$\times$10$^{-5}$ counts s$^{-1}$ arcmin$^{-2}$.
To this background value we must add the photons in the wings of the
PSF.  Sources 3, 5, 9, and 10 contribute $>$90\% of the total counts
in the 0.5-2.0 keV band.  If we assume these counts raise the
background uniformly, then an additional $\sim$2.2$\times$10$^{-5}$
counts s$^{-1}$ arcmin$^{-2}$ exists.  The ratio of the detected
diffuse emission to the sum of the backgrounds is $\sim$1.2.

We set a limit on the luminosity of additional point sources of
$\sim$5$\times$10$^{37}$ erg s$^{-1}$, about 25\% lower than the limit
from the PSPC data.  Either the diffuse emission is truly diffuse or a
gap exists, at about a few 10$^{37}$ erg s$^{-1}$, in the point source
luminosity function for this galaxy.  The definitive properties of the
diffuse emission must await observations with the {\it Chandra X-ray
Observatory} or {\it XMM}.

\subsection{The MF16 SNR Complex} \label{SeceSNR}

A sensitive search for emission lines from the hot gas in the spectrum
of MF16 constituted one of the prime purposes for the {\it ASCA}
observation.  We extracted the SIS source counts in a region 8$'$ in
radius.  We only used the SIS data because of that instrument's
superior spectral resolution.  For the background, we extracted the
counts from all pixels outside of the source aperture.  We tested the
background subtraction by extracting the counts from a ``blank sky''
observation\footnote{``Blank sky'' observations are available from the
{\it ASCA} GOF at NASA-GSFC.}  using the source aperture.

Extracted SIS data were then used to construct a model fit.  Because
the low energy calibration of the SIS is inaccurate below $\sim$0.5
keV \citep{Do97} while the {\it ROSAT} PSPC is sensitive to the column
density, N$_{\rm H}$, we combined the {\it ROSAT} PSPC and the SIS
data.  MF16's evolution should be sufficiently slow that little
spectral evolution will occur during the $\sim$2.5 year gap between
observations.  The {\it ROSAT} PSPC spectrum was extracted using an
aperture of 1$'$; the background was defined by an aperture 8$'$ due
west and well outside the galaxy's detected diffuse emission
structure.  We temporarily ignored data near the positions of expected
lines (e.g., Fe K${\alpha}$, Fe L, Si K${\alpha}$) and used the
remaining SIS+PSPC data to define the continuum.  Once we had a
successful continuum fit, we used all the data to search for emission
lines in the SIS spectrum by adding one or more gaussians to represent
the line(s).  The fit results are listed in Table~\ref{sum_Fit}.

The resulting continuum contours after fitting the PSPC+SIS spectrum
with an absorbed power law are shown in Figure~\ref{GIS_cont}.  From
the dust maps of \cite{SFD98}, the value of E$_{\rm B-V}$ in the
direction of NGC 6946 is 0.342.  Using the column density-E$_{\rm
B-V}$ relation of \cite{PS95}, that value of E$_{\rm B-V}$ converts to
N$_{\rm H}$ $\sim$1.8$\times$10$^{21}$ cm$^{-2}$.  The column density
derived from the model fit to the {\it ASCA} data is
$\sim$2$\pm$0.2$\times$10$^{21}$ cm$^{-2}$.  The measured E$_{\rm
{B-V}}$ toward the MF16 complex is 0.65 (BFS00) which implies patchy
extinction on an unresolved spatial scale.  Very likely, the X-ray
column is an "effective" column that arises from the area-weighted
average of spatially-variable extinction.  As a test of our
interpretation, when we fix the column at the BFS00 value, the
resulting fit is very poor with ${\Delta}{\chi}^2/{\nu}$ $\sim$1.5.

The best single component fit to the continuum yields a power law
index of $\sim$2.5.  The unavoidable presence of the neighboring
source H8 in the combined spectrum may artificially lower the power
law index by inputting more energy above $\sim$5 keV than the SNR
alone.  Nonetheless, from our fit we estimate an unabsorbed 0.5--2.0
keV luminosity of $\sim$2$\times$10$^{40}$ erg s$^{-1}$ and the
2.0--10.0 keV luminosity of $\sim$7$\times$10$^{39}$ erg s$^{-1}$.

We see distinct residuals near 0.8--0.9 keV (Figure~\ref{SIS_spec}).
The apparent line which is best fit by a width (${\sigma}$) of
$\sim$0.1 keV centered at 0.91 keV (Figure~\ref{SIS_line}).  The
equivalent width of the line is 157$\pm$50 eV.  If we interpret the
line as a ${\delta}$-function at the measured energy, it corresponds
to Ne IX at either 0.915 or 0.922 keV.  Conversely, the line might
represent a blend of Fe L-shell emission from Fe XIX and Fe XX
\citep{Kall96}.  However, unless the Fe abundance is enhanced, high
ionization stages of Fe L-shell emission are accompanied by
K${\alpha}$ lines of medium-Z elements (e.g., Ne, Mg).  Unfortunately,
we have no information regarding the Fe abundance from the {\it ASCA}
data.  The poor signal-to-noise above $\sim$5 keV places only a weak
limit on the pressence of an line in the 6.4-6.7 keV band (no line
$>$1 keV equivalent width), so we can not deduce any information about
the presence of Fe L based on the presence or absence of Fe K.

Two alternative, equally valid fits use either dual bremsstrahlung or
dual Raymond-Smith components.  We followed the same fitting procedure
(i.e., first PSPC+SIS continuum, then line search).  In neither case
do we detect emission lines in the model fits.  The fitted
bremsstrahlung temperatures are 0.21$^{+0.02}_{-0.04}$ and
3.0$^{+0.4}_{-0.6}$ keV, respectively.  The fitted data, using the
Raymond-Smith model, appears in Figure~\ref{SIS_Dual_spec} and the
parameter contours are shown in Figure~\ref{SIS_Dual_cont_ray}.  The
fitted Raymond-Smith temperatures are 0.81$^{+0.04}_{-0.05}$ keV and
4.2$^{+0.6}_{-0.4}$ keV.  The unabsorbed 0.5--2.0 keV luminosity is
$\sim$9.6$\times$10$^{39}$ erg s$^{-1}$ and the 2--10 keV luminosity
is $\sim$1.1$\times$10$^{40}$ erg s$^{-1}$ which are about a factor of
2 lower and 1.6 higher, respectively, than the computed luminosities
from the power law model.  The lower flux in the 0.5--2.0 keV band
comes from the improved fit to the low-energy channels which reduces
the column density.

Although the {\it ASCA} spectra do not have sufficient signal-to-noise
to constrain the abundance values, we can test their effects.  For an
abundance of 0.1 solar, the component temperatures lie well within the
errors of the original fit.  No abundance peculiarities were detected
in the optical spectrum either, although the strength of the [N II]
6584 {\AA} line relative to H${\alpha}$ was considerably stronger in
MF16 than in any other SNR in NGC 6946 and indicates possible
enrichment of nitrogen (BFS00).

As an aid to interpreting the observed X-ray data on MF16, we used HST
optical images (BFS00) which provide a wide-field view of the region.
Within a square $\sim$30$''$ region with its eastern edge fixed on
MF16, potential X-ray contributors include the SNR MF15 ($\sim$15$''$
to the NW), an apparent OB association ($\sim$10$''$ to the W), a
possible X-ray binary (XRB, 30$''$ to the SW), and MF16 itself.  SNR
MF15 is about an order of magnitude less bright than MF16 in the
optical and may be similarly faint in the X-ray.  Also, because the OB
association shows no evidence of abnormal brightness in the {\it HST}
image, we assume its X-ray emission is the product of the emission
from an average OB star and the number of member stars; that product
is $\sim$10$^{34-36}$ erg s$^{-1}$ (e.g., \citealt{GC94}), well below
the luminosity of MF16.  The possible XRB undoubtedly contaminates the
spectrum, particularly at high energies, but with a luminosity ratio
of $\sim$15 to 1, the spectrum will not be dominated by the XRB's
emission below $\sim$3 keV.

Determining the precise physical locations of the observed X-rays
coming from MF16 is not possible from the present
low-spatial-resolution data sets.  However, several sources are
likely: each of the SNR shells or the optically bright,
crescent-shaped interaction region where two of the SNRs are
apparently colliding (BFS00).  Each source's emission, unfortunately,
cannot be uniquely deconvolved from the observed spectrum and two
equally valid models fit the MF16 spectrum: a power law continuum plus
emission line or dual thermal components consisting of bremsstrahlung
or Raymond-Smith spectra.  The plausibility of a thermal model is
questionable on the grounds of a lack of thermal line emission, yet
appears more physically likely given that other SNRs possess thermal
components with temperatures of $\sim$0.2 keV (e.g., the Cygnus Loop,
\citealt{Miy94}).  Therefore, in the following discussion, we adopt
the dual Raymond-Smith model.

If we treat the interaction of the two shocks as approximately
equivalent to a shock impacting a wall (as in the Cygnus Loop;
\citealt{Lev97, HRB94}), then the cooler component of the {\it ASCA}
spectrum should describe the summed emission of both SNRs, leaving the
hard spectral component to describe the interaction region.  This
component assignment has a precedent.  Simulations (e.g.,
\citealt{PS97}) and {\it ASCA} observations \citep{Mae99} of colliding
winds in O star binaries show the harder component ($\sim$2--3 keV) is
emitted by the interaction region.  The emission measure ($EM \sim n^2
V$, for density $n$ and volume $V$) is directly related to the
normalization of the Raymond-Smith component.  Assuming two spherical
SNRs of diameters of $\sim$8 and $\sim$20 pc (BFS00) each contributing
half the total, we estimate electron densities of $\sim$35 and
$\sim$10 cm$^{-3}$, respectively, in good agreement with the pre-shock
density values inferred from the post-shock, optical [S II]
$\lambda\lambda$6716,6731 line ratio (BFS00).

For the optically bright interaction crescent region, we estimate an
upper limit to the electron density assuming that the hard spectral
component is not strongly contaminated by emission from other nearby
sources.  Model simulations of colliding SNRs show such interactions
will generate a ring-like contact zone which then expands outward as
the shells merge (e.g., \citealt{Ike78, VC95}).  Assuming the
interaction region is also a torus, the fitted emission measure and
the toroid volume yield a density of $\sim$100 cm$^{-3}$.
Contamination of the hard component will lower this estimate.  At the
interface, we expect the thermal and ram pressures of each SNR to be
approximately equal.  Since the density ratio is $\sim$3--4, the shock
velocity ratio must be $\sim$1.7--2.  Spatially-resolved X-ray
spectroscopy will provide valuable constraints on the relative
strengths of the two shocks.

Models of colliding SNRs assume a homogeneous circum-progenitor medium
\citep{Ike78,VC95}.  If such a medium described the MF16 region, the
SNRs would eventually merge to one remnant (e.g., \citealt{Ike78}).
However, we have uncovered evidence for an inhomogeneous medium.
Furthermore, the {\it HST} data suggest the possibility of a cavity
explosion (BFS00), further complicating the dynamics and clouding our
understanding of the origin and evolution of the SNR complex.  Given
the already high optical, radio, and X-ray luminosity of this
remarkable SNR complex, such a merger could be related to especially
large, energetic SNRs (hypernovae remnants) recently identified in
other galaxies (M101; \citet{Wng99}).  In any case, further study of
this extraordinarily luminous X-ray region should help shed new
insights on the properties of SNR shock emission under high density
situations.

\section{Summary}

We have presented the data from an {\it ASCA} and a {\it ROSAT} HRI
observation of the spiral galaxy NGC 6946.  The {\it ASCA} image
represents the first look at this galaxy above 2 keV.  The {\it ASCA}
image contains evidence for diffuse emission in the inner $\sim$4$'$
extending to energies $>$2 keV.  Fourteen point-like sources are
detected in the HRI observation.  One source corresponds to the very
luminous SNR complex uncovered in the PSPC data.  The HRI resolves a
point source $\sim$30$''$ west of the SNR complex that could not be
detected in the PSPC data.  The luminosity of the point source is
$\sim$1/15 of the SNR complex.  None of the point sources shows
evidence of variability.

Two possible spectral fits to the {\it ASCA} spectrum of the SNR
complex provide contrasting interpretations.  The spectrum can be fit
either by a power law plus an emission line or by dual Raymond-Smith
thermal plasma models with differing temperatures.  We argue that the
dual thermal models provide a physically plausible interpretation.

\acknowledgments

This research was supported by grant NAG5-4015 to SAO from the {\it
ASCA} Guest Observer Program.  EMS thanks John Raymond for a valuable
conversation on possible X-ray spectra from colliding SNRs.

\newpage

\begin{table*} 
\begin{center}
\caption{Observation Summary} \label{sum_Xobs}
\begin{tabular}{llllrl}
Satellite & Instruments & Dates & JD & Exposure Time & Sequence \cr \hline
 {\it ROSAT} & PSPC & 1992 June 16--21 & 48789--94 & 36.7 ksec & rp600272 \\
 {\it ROSAT} & HRI  & 1994 May 14--29 & 49486--501 & 60.3 ksec & rh600501n00 \\
 {\it ASCA}  & SIS, GIS & 1994 Dec 28 to & 49714--19 & 83.7 ksec & 53040000 \\
             &          &  1995 Jan 2    &       &           &      \\ \hline
\end{tabular}
\end{center}
\end{table*}

\newpage

\begin{deluxetable}{cllrrrrrr} 
\tablecaption{{\it ROSAT} HRI Point Sources within 8' of the Center of NGC 6946 \label{pt_srcs}}
\tablehead{
\colhead{} &
\colhead{} &
\colhead{} &
\colhead{Counts} &
\colhead{} &
\colhead{} &
\colhead{} &
\colhead{} &
\colhead{} \\
\colhead{No.} &
\colhead{RA (J2000)\tablenotemark{a}} &
\colhead{Dec (J2000)\tablenotemark{a}} &
\colhead{$\pm$Error} &
\colhead{S/N} &
\colhead{Rate\tablenotemark{b}} &
\colhead{Flux\tablenotemark{c}} &
\colhead{L$_{\rm X}$\tablenotemark{d}} &
\colhead{Notes\tablenotemark{e}}}
\startdata
 1 & 20:34:25.9 & +60:09:05.4 &  64$\pm$10.7 &  7.8 &  10.7$\pm$1.8 &  2.8$\pm$0.7 &   8.7$\pm$2.1 & x \\
 2 & 20:34:34.5 & +60:10:30.4 &  45$\pm$10.4 &  4.7 &   5.9$\pm$2.3 &  1.7$\pm$0.6 &   5.3$\pm$2.0 & ... \\
 3 & 20:34:36.6 & +60:09:29.9 & 138$\pm$14.5 & 14.9 &  23.3$\pm$2.4 &  6.6$\pm$0.8 &  20.5$\pm$2.5 & x, o?, r \\
 4 & 20:34:48.7 & +60:05:48.5 &  42$\pm$10.2 &  4.8 &   6.9$\pm$2.3 &  2.3$\pm$0.7 &   7.1$\pm$2.0 & x \\
 5 & 20:34:52.4 & +60:09:11.2 & 167$\pm$18.7 & 11.5 &  27.9$\pm$3.1 &  9.3$\pm$0.9 &  28.8$\pm$2.8 & x,r? \\
 6 & 20:34:56.8 & +60:08:33.7 &  65$\pm$14.6 &  4.8 &  10.9$\pm$2.4 &  3.1$\pm$0.7 &   9.6$\pm$2.2 & x, r \\
 7 & 20:34:57.8 & +60:09:47.3 &  57$\pm$14.3 &  4.3 &   9.6$\pm$2.4 &  2.7$\pm$0.7 &   8.4$\pm$2.1 & ... \\
 8 & 20:35:00.6 & +60:11:29.4 &  52$\pm$10.1 &  4.2 &   8.7$\pm$1.7 &  2.5$\pm$0.5 &   7.8$\pm$1.5 & ... \\
 9 & 20:35:00.3 & +60:09:05.8 & 196$\pm$15.9 & 20.3 &  31.1$\pm$2.6 &  9.7$\pm$0.9 &  30.0$\pm$2.8 & x \\
10 & 20:35:00.7 & +60:11:29.4 & 827$\pm$31.4 & 61.3 & 138.2$\pm$5.2 & 38.2$\pm$1.4 & 118.4$\pm$4.5 & x, o, r \\
11 & 20:35:01.5 & +60:10:04.4 &  45$\pm$11.3 &  4.4 &   7.6$\pm$1.9 &  2.8$\pm$0.7 &   8.7$\pm$2.1 & x \\
12 & 20:35:11.7 & +60:07:30.0 &  34$\pm$ 9.6 &  4.0 &   5.7$\pm$1.6 &  1.5$\pm$0.6 &   4.6$\pm$1.9 & ... \\
13 & 20:35:18.0 & +60:10:54.1 &  41$\pm$10.2 &  4.5 &   6.9$\pm$1.7 &  1.7$\pm$0.6 &   6.3$\pm$2.4 & ... \\ 
14 & 20:35:20.7 & +60:10:16.6 &  52$\pm$12.1 &  4.1 &   8.7$\pm$2.4 &  2.5$\pm$0.7 &   7.8$\pm$2.1 & ... \\
\enddata
\tablenotetext{a}{Positions accurate to ${\sim}{\pm}$1$''$.}

\tablenotetext{b}{Units = 10$^{-4}$ counts s$^{-1}$}

\tablenotetext{c}{Flux, in units of 10$^{-14}$ erg s$^{-1}$
cm$^{-2}$, calculated by adopting a 1 keV thermal bremsstrahlung
spectrum absorbed by a column of 2.5$\times$10$^{21}$ cm$^{-2}$, so 1
count s$^{-1}$ $\sim$2.8$\times$10$^{-10}$ erg s$^{-1}$ cm$^{-2}$ in
the 0.5--2.0 keV band}

\tablenotetext{d}{Units = 10$^{37}$ erg s$^{-1}$}

\tablenotetext{e}{``Notes'' Column: The counterparts in other papers are
indicated by a letter.  Each letter is identified below.  Question
marks over an equals sign denote a possibly dubious identification.
  {\bf x} = X-ray: PSPC data (Schlegel 1994a): 1 = PSPC-8; 3 = PSPC-2; 4 =
PSPC-9; 5 = PSPC-3; 6 = PSPC-6; 9 = PSPC-4; 10 = PSPC-1; 11 = PSPC-7;
sources 2, 7, 12, 13, and 14 are weak, but visible in the PSPC data;
source 8 is unresolved in the PSPC data.
  {\bf r} = Radio: Lacey, Duric, \& Goss (1997) (LDG97): 3 = LDG97-22;
5 ${? \over =}$ LDG97-56; 6 = LDG97-80; 10 = LDG97-85.
  {\bf o} = optical: Matonick \& Fesen (1997) (MF97): 10 = MF97-16; 3
${? \over =}$ MF97-4.}

\end{deluxetable}

\newpage

\begin{deluxetable}{llllrrr}
\tablecaption{Summary of Spectral Fitting\tablenotemark{a}~~~for the SNR MF16 \label{sum_Fit}}
\tablehead{
\colhead{} &
\colhead{N$_{\rm H}$} &
\colhead{First} &
\colhead{Second} &
\colhead{Mean SIS} &
\colhead{Ratio of} &
\colhead{Overall} \\
\colhead{Model} &
\colhead{$\times$10$^{21}$ cm$^{-2}$} &
\colhead{Parameter\tablenotemark{b}} &
\colhead{Parameter\tablenotemark{b}} &
\colhead{Norm} &
\colhead{Norms (1:2)} &
\colhead{${\chi}^2$/${\nu}$}
}
\startdata
Power Law              & 1.88$^{+0.32}_{-0.28}$ & 2.34$^{+0.14}_{-0.11}$ & $\cdots$                    & 4.4(--4) & 2.1  & 0.98 \\
{\indent} + Gaussian   & & 0.90$^{+0.05}_{-0.04}$ & 0.10$^{+0.05}_{-0.04}$ & 2.2(--4) & $\cdots$  & $\cdots$ \\
Brems-1                & 4.96$^{+0.42}_{-0.67}$ & 0.22$^{+0.04}_{-0.07}$ & $\cdots$                    & 7.7(--2) & 148  & 0.99 \\
{\indent} + Brems-2    & & 3.0$^{+0.7}_{-1.0}$    & $\cdots$                    & 5.2(--4) & $\cdots$  & $\cdots$ \\
Raymond-1              & 0.83$^{+0.22}_{-0.14}$ & 0.83$^{+0.02}_{-0.08}$ & 1.0                    & 1.0(--4) & 0.14 & 1.01 \\
{\indent} + Raymond-2  & & 4.1$^{+0.8}_{-0.2}$    & 1.0                    & 7.2(--4) & $\cdots$  & $\cdots$ \\
Low-Z Ray-1            & 0.99$^{+0.14}_{-0.12}$ & 0.86$^{+0.05}_{-0.08}$ & 0.5                    & 2.1(--4) & 0.19 & 1.03 \\
{\indent} + Low-Z Ray-2& & 3.2$^{+0.9}_{-0.6}$    & 0.5                    & 1.1(--3) & $\cdots$  & $\cdots$ \\
\enddata
\tablenotetext{a}{All of the fits were carried out first using the
PSPC + SIS data with regions near lines screened out of the fit and
then searching for lines using the all the data.}

\tablenotetext{b}{Parameters are defined as:
{\bf Bremsstrahlung}:  first: temperature (keV), second: none;
{\bf Gaussian}:  first: line position (keV), second: line width (keV);
{\bf Power Law}:  first: power law index, second: none;
{\bf Raymond-Smith}:  first: temperature (keV), second: abundance (fixed during the fit)
}
\end{deluxetable}

\newpage

\begin{figure}
\caption{The {\it ROSAT} HRI image of NGC 6946.  North is up, east is
left. The panel is 8$'$ on a side.  The data have been screened and
adaptively smoothed using a smoothing scale of 10 counts.  The point
sources from Table~\ref{pt_srcs} are identified.}  \label{HRI_fig}
\end{figure}

\begin{figure}
\caption{The HRI image, contoured, on top of a digitized POSS image of NGC
6946.  The images are 8$'$ on a side with North up, East left.  The figures show the upper (left) and lower (right) ranges of the image scaling.}
\label{HRI_opt}
\end{figure}

\begin{figure}
\caption{A montage of deconvolved {\it ASCA} images of NGC 6946.  The
energy band of the data in the individual panels increases to the
right and to the bottom.  The energies included in the panels are:
0.5-1.0, 1.0-1.5, 1.5-2.5, 2.5-3.5, 3.5-4.5, and 4.5-9.5 keV.  The
bottom right panel is the sum of all of the bands. North is up, east
is left.  Each panel is 8$'$ on a side.  The `+' sign in the 1.5-2.5
and 4.5-9.5 panels indicates the centroid position of MF16 from the
1.0-1.5 keV panel.  Note how the centroid of emission shifts westward
in the higher energy panel.} \label{ASCA_montage}
\end{figure}

\begin{figure}
\caption{The HRI image blurred with the {\it ASCA} PSF and compared
directly to the background-subtracted {\it ASCA} image.  The top left
panel shows the screened HRI data, extracted to match the centering
and scale, but {\it not} the rotation angle, of the {\it ASCA} data.
The top right panel shows the HRI data blurred by the {\it ASCA} PSF
for E $<$ 2.5 keV.  The bottom right panel shows the {\it ASCA} SIS
data filtered to include only events with E $<$ 2.5 keV.  Note the
slight rotation of the {\it ASCA} image ($\sim$20 degrees,
counterclockwise).  North is up, east is left. Each box is 8$'$ on a
side.} \label{hri_blur}
\end{figure}

\begin{figure}
\caption{The contours on the fitted continuum parameters for the MF16
region from the PSPC+GIS spectral data using a power law to fit the
continuum.}
\label{GIS_cont}
\end{figure}

\begin{figure}
\caption{The PSPC+SIS spectrum of the MF16 region.  For clarity, only
the PSPC spectrum and one SIS spectrum are shown (top).  The
normalization of the line model has been set to zero to enhance its
visibility. The continua were fit using a power law.  (bottom) The
contributions to the remaining chi-squared of the fit.}
\label{SIS_spec}
\end{figure}

\begin{figure}
\caption{The contours on the fitted emission line parameters from the
PSPC+SIS fit using a power law for the continuum.} \label{SIS_line}
\end{figure}

\begin{figure}
\caption{The spectral fit using the dual Raymond-Smith model.  Only
the PSPC spectrum and one SIS spectrum are plotted for clarity (top).
(bottom) The contributions to the remaining chi-squared of the
fit.}\label{SIS_Dual_spec}
\end{figure}

\begin{figure}
\caption{The contours on the fitted continuum parameters using the
dual Raymond-Smith model applied to the PSPC+SIS
data.}\label{SIS_Dual_cont_ray}
\end{figure}

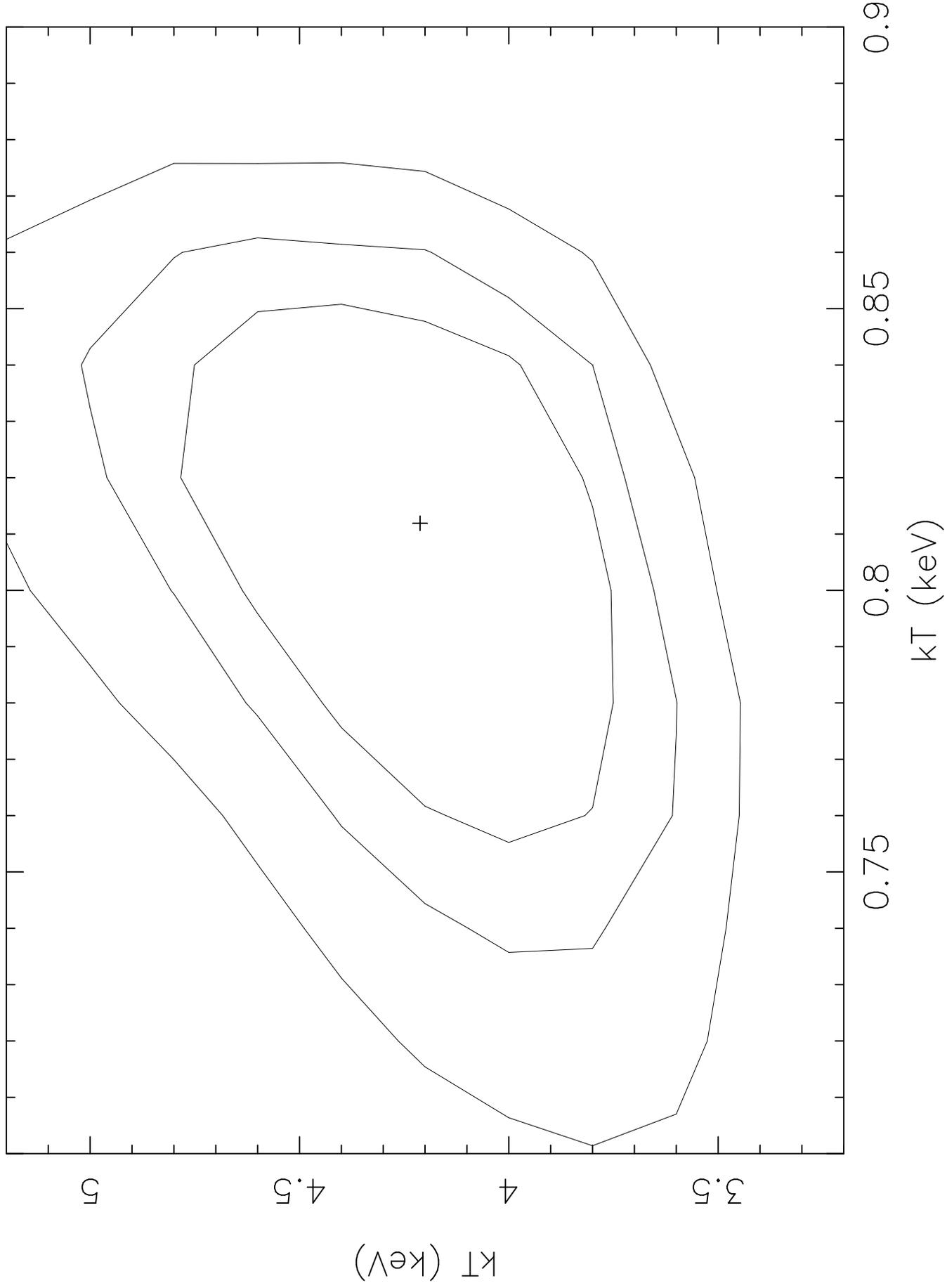
\end{document}